\documentclass[aps,prl,preprint,tightenlines,superscriptaddress,showpacs,byrevtex]{revtex4}
\usepackage{graphicx} 
\usepackage{dcolumn}  

\graphicspath{{ps}}

\newcommand{\ee}{e^{+}e^{-}}
\newcommand{\leplep}{\ell^{+}\ell^{-}}
\newcommand{\jp}{J/\psi}

\newcommand{\psip}{\psi(2S)}

\newcommand{\pipi}{\pi^{+}\pi^{-}}
\newcommand{\piz}{\pi^{0}}
\newcommand{\bbar}{B\bar{B}}

\newcommand{\Mbc}{M_{\rm bc}}
\newcommand{\DE}{\Delta E}

\newcommand{\rt}{\rightarrow}

\newcommand{\etal}{\em et al.}

\begin{document}


\preprint{\vbox{ \hbox{   }
                   \hbox{BELLE-CONF-0540}
                   \hbox{LP2005-175} 
}}

\title{ \quad\\[0.5cm] Evidence for $X(3872)\rt\gamma\jp$ and 
the sub-threshold decay $X(3872)\rt\omega\jp$  }

\affiliation{Aomori University, Aomori}
\affiliation{Budker Institute of Nuclear Physics, Novosibirsk}
\affiliation{Chiba University, Chiba}
\affiliation{Chonnam National University, Kwangju}
\affiliation{University of Cincinnati, Cincinnati, Ohio 45221}
\affiliation{University of Frankfurt, Frankfurt}
\affiliation{Gyeongsang National University, Chinju}
\affiliation{University of Hawaii, Honolulu, Hawaii 96822}
\affiliation{High Energy Accelerator Research Organization (KEK), Tsukuba}
\affiliation{Hiroshima Institute of Technology, Hiroshima}
\affiliation{Institute of High Energy Physics, Chinese Academy of Sciences, Beijing}
\affiliation{Institute of High Energy Physics, Vienna}
\affiliation{Institute for Theoretical and Experimental Physics, Moscow}
\affiliation{J. Stefan Institute, Ljubljana}
\affiliation{Kanagawa University, Yokohama}
\affiliation{Korea University, Seoul}
\affiliation{Kyoto University, Kyoto}
\affiliation{Kyungpook National University, Taegu}
\affiliation{Swiss Federal Institute of Technology of Lausanne, EPFL, Lausanne}
\affiliation{University of Ljubljana, Ljubljana}
\affiliation{University of Maribor, Maribor}
\affiliation{University of Melbourne, Victoria}
\affiliation{Nagoya University, Nagoya}
\affiliation{Nara Women's University, Nara}
\affiliation{National Central University, Chung-li}
\affiliation{National Kaohsiung Normal University, Kaohsiung}
\affiliation{National United University, Miao Li}
\affiliation{Department of Physics, National Taiwan University, Taipei}
\affiliation{H. Niewodniczanski Institute of Nuclear Physics, Krakow}
\affiliation{Nippon Dental University, Niigata}
\affiliation{Niigata University, Niigata}
\affiliation{Nova Gorica Polytechnic, Nova Gorica}
\affiliation{Osaka City University, Osaka}
\affiliation{Osaka University, Osaka}
\affiliation{Panjab University, Chandigarh}
\affiliation{Peking University, Beijing}
\affiliation{Princeton University, Princeton, New Jersey 08544}
\affiliation{RIKEN BNL Research Center, Upton, New York 11973}
\affiliation{Saga University, Saga}
\affiliation{University of Science and Technology of China, Hefei}
\affiliation{Seoul National University, Seoul}
\affiliation{Shinshu University, Nagano}
\affiliation{Sungkyunkwan University, Suwon}
\affiliation{University of Sydney, Sydney NSW}
\affiliation{Tata Institute of Fundamental Research, Bombay}
\affiliation{Toho University, Funabashi}
\affiliation{Tohoku Gakuin University, Tagajo}
\affiliation{Tohoku University, Sendai}
\affiliation{Department of Physics, University of Tokyo, Tokyo}
\affiliation{Tokyo Institute of Technology, Tokyo}
\affiliation{Tokyo Metropolitan University, Tokyo}
\affiliation{Tokyo University of Agriculture and Technology, Tokyo}
\affiliation{Toyama National College of Maritime Technology, Toyama}
\affiliation{University of Tsukuba, Tsukuba}
\affiliation{Utkal University, Bhubaneswer}
\affiliation{Virginia Polytechnic Institute and State University, Blacksburg, Virginia 24061}
\affiliation{Yonsei University, Seoul}
  \author{K.~Abe}\affiliation{High Energy Accelerator Research Organization (KEK), Tsukuba} 
  \author{K.~Abe}\affiliation{Tohoku Gakuin University, Tagajo} 
  \author{I.~Adachi}\affiliation{High Energy Accelerator Research Organization (KEK), Tsukuba} 
  \author{H.~Aihara}\affiliation{Department of Physics, University of Tokyo, Tokyo} 
  \author{K.~Aoki}\affiliation{Nagoya University, Nagoya} 
  \author{K.~Arinstein}\affiliation{Budker Institute of Nuclear Physics, Novosibirsk} 
  \author{Y.~Asano}\affiliation{University of Tsukuba, Tsukuba} 
  \author{T.~Aso}\affiliation{Toyama National College of Maritime Technology, Toyama} 
  \author{V.~Aulchenko}\affiliation{Budker Institute of Nuclear Physics, Novosibirsk} 
  \author{T.~Aushev}\affiliation{Institute for Theoretical and Experimental Physics, Moscow} 
% ...
\collaboration{The Belle Collaboration}

\noaffiliation

\begin{abstract}
We report evidence for the decay modes $X(3872)\rt\gamma\jp$ and
$X(3872)\rt\pipi\piz\jp$.
In the latter, the $\pipi\piz$ invariant mass distribution has a
strong peak between 750~MeV and the kinematic limit of 775~MeV,
suggesting that the process is dominated by the sub-threshold
decay $X\rt\omega\jp$.  These results establish the charge-conjugation
parity of the $X(3872)$ as $C=+1$.
The results are based on a study of 
$X(3872)$ mesons produced via exclusive $B^{\mp} \rt K^{\mp} X(3872)$
decays in a 256~fb$^{-1}$ data sample
collected at the $\Upsilon(4S)$ resonance in the Belle detector at 
the KEKB collider.   The
$X(3872)\rt\gamma\jp$ signal and
$\pipi\piz\jp$ signal each has a statistical
significance that is greater than $4\sigma$.
\end{abstract}

\pacs{14.40.Gx, 12.39.Mk, 13.20.He}

\maketitle


{\renewcommand{\thefootnote}{\fnsymbol{footnote}}}
\setcounter{footnote}{0}

The $X(3872)$ was discovered by Belle as a 
narrow $\pipi\jp$ mass peak in exclusive 
$B^-\rt K^-\pipi\jp$ decays~\cite{skchoi_x3872,conj}.  
The observed mass and the narrow width are not compatible
with expectations for any of the as-yet unobserved 
charmonium states~\cite{solsen_krakow}.  Moreover, the
$\pipi$ invariant mass distribution peaks
near the upper kinematic limit of $M(\pipi)=775$~MeV, as expected
for dipions that originate from $\rho\rt\pipi$ decays.  
Charmonium decays to $\rho\jp$ final states violate isospin and are 
expected to be  suppressed.  The $X(3872)$ and its above-listed 
properties were confirmed by other experiments~\cite{X_BaBar}.

The $X(3872)$ mass ($3871.9\pm 0.5$~MeV~\cite{X_mass}) 
is within errors of the  $D^0\bar{D^{0*}}$ threshold 
($3871.3\pm 1.0$~MeV~\cite{PDG});  the difference is 
$0.6\pm 1.1$~MeV. This has led to speculation that the 
$X$ might be a molecule-like $D^0\bar{D}^{0*}$ bound 
state~\cite{tornqvist,swanson_1,molecule}. According to 
Ref.~\cite{tornqvist}, the preferred quantum numbers for 
such a bound state would be either $J^{PC} = 0^{-+}$ or 
$1^{++}$. The decay of a $C=+1$ state to $\pipi\jp$ would 
proceed via an isovector $\rho^0\jp$ intermediate state and 
produce a $\pipi$ mass spectrum that is concentrated
at high masses, as is observed.  In this meson-meson 
bound state interpretation, the close proximity of the 
$X$ mass to $D^0\bar{D}^{0*}$ threshold compared to the 
$D^+ \bar{D}^{*-}$-$D^0\bar{D}^{0*}$ mass splitting
of 8.1~MeV produces a strong isospin violation.

Swanson proposed a dynamical model for the $X(3872)$ as a 
$D^0\bar{D}^{0*}$ hadronic resonance~\cite{swanson_1}.  
In this model, $J^{PC}=1^{++}$ is strongly favored and
the wave function has, in addition to $D^0\bar{D}^{0*}$,
an appreciable admixture of $\omega\jp$ plus
a small $\rho\jp$ component.  The latter produces
the $\pipi\jp$ decays that have been observed; the
former gives rise to $\pipi\piz\jp$ decays via a
virtual $\omega$ that is enhanced because of
the large $\omega\jp$ component to the wavefunction.
Swanson's model predicts that $X(3872)\rt \pipi\piz\jp$ 
decays should occur at about half the rate for 
$\pipi\jp$ and with a $\pipi\piz$ invariant mass 
spectrum that peaks near the upper kinematic boundary 
of 775~MeV (7.5~MeV below the $\omega$ peak).  In a 
subsequent paper~\cite{swanson_2}, Swanson emphasized 
the importance  of searching for the radiative decay 
$X(3872)\rt\gamma\jp$. The observation of this decay 
mode would unambiguously establish the charge-conjugation 
parity of the $X(3872)$ as $C=+1$.

In this Letter we report on a search for $B\rt K\gamma\jp$
and $K\pipi\piz\jp$ decays in a 275~million $\bbar$ event sample 
collected in the Belle detector at the KEKB energy-asymmetric 
$\ee$ collider.  The data were accumulated at a center-of-mass 
system (cms) energy of $\sqrt{s} = 10.58$~GeV, corresponding to 
the mass of the $\Upsilon(4S)$ resonance.  KEKB is described 
in detail in ref.~\cite{KEKB}.

The Belle detector is a large-solid-angle magnetic 
spectrometer  that consists of a three-layer silicon vertex 
detector, a 50-layer cylindrical drift chamber (CDC), an 
array of aerogel threshold Cherenkov counters (ACC),  a 
barrel-like arrangement of time-of-flight  scintillation 
counters (TOF), and an electromagnetic calorimeter
(ECL) comprised of CsI(Tl) crystals  located inside
a superconducting solenoid coil that provides a 1.5~T
magnetic field.  An iron flux-return located outside of 
the coil is instrumented to detect $K_L$ mesons and to 
identify muons (KLM).  The detector is described in detail 
elsewhere~\cite{Belle}.

For the $X(3872)\rt\gamma\jp$ study we select events that contain a 
$\jp$, either a charged or neutral kaon and a $\gamma$ with 
laboratory-frame energy greater than 40~MeV.  We use
the $\jp$ and kaon selection criteria described in 
refs.~\cite{skchoi_x3872} and~\cite{skchoi_y3940}.  The $\gamma$
candidate is rejected if, when combined with any other photon in the
event, fits a $\pi^0\rt\gamma\gamma$ hypothesis with
$\chi^2 <4$. 
To reduce the level of $\ee\rt q\bar{q}$ ($q=u,d,s~{\rm or}~c$-quark)
continuum events in the sample,
we also require  $R_2 < 0.4$, where $R_2$ is the normalized
Fox-Wolfram moment~\cite{fox}, and $|\cos\theta_B|<0.8$, where
$\theta_B$ is the polar angle of the $B$-meson direction
in the cms.

Candidate $B \rt K\gamma\jp$ mesons are identified by the 
energy difference  $\DE\equiv E_B^{\rm cms} - E_{\rm beam}^{\rm cms}$
and the beam-energy constrained mass 
$\Mbc\equiv\sqrt{(E_{\rm beam}^{\rm cms})^2-(p_B^{\rm cms})^2}$,
where $E_{\rm beam}^{\rm cms}$ is the cms beam
energy, and $E_B^{\rm cms}$ and $p_B^{\rm cms}$ are the cms 
energy and momentum of the $K\gamma\jp$ combination.  We select 
events with $M_{bc}>5.20$~GeV and $|\DE|<0.2$~GeV and among 
these define a signal region $5.2745~{\rm GeV} < \Mbc <5.2855$~GeV
and $|\DE |< $ 0.035 GeV, which correspond to $\simeq \pm 2\sigma$ 
from the central values for both variables.

There is prominent peak near $3510$~MeV in the
$M(\gamma\jp)$ distribution~\cite{m3pijpsi}, due to 
$B\rt K\chi_{c1}$;  $\chi_{c1}\rt\gamma\jp$. 
This is used as a calibration reaction to
determine $\Mbc$, $\DE$ and $M(\gamma\jp)$
resolutions and shifts in the nominal peak
positions.

\begin{figure}[htb]
\includegraphics[width=0.6\textwidth]{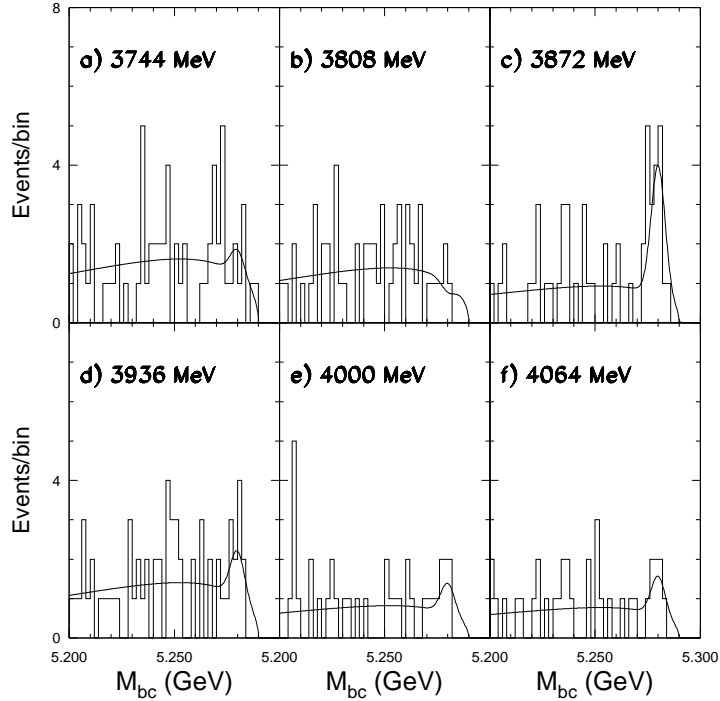}
\caption{\label{fig:mb_gamjpsi_6box}
The $\Mbc$ distributions
for 64-MeV-wide $M(\gamma\jp)$ mass bins in the $X(3872)$ mass region.
The curves are the results of fits described in the text.
}
\label{}
\end{figure}

Figures~\ref{fig:mb_gamjpsi_6box}(a) through (f)
show the $\Mbc$ distributions for six 64-MeV-wide
$M(\gamma\jp)$ mass bins in the $X(3872)$ region.
The 64~MeV bin-width corresponds
to $\pm 2.5\sigma$, where $\sigma$ is the $M(\gamma\jp)$
mass resolution determined by scaling the measured
width of the $\chi_{c1}$ peak to the $X(3872)$ region
using a MC-determined factor.  The upper right-hand 
panel of Fig.~\ref{fig:mb_gamjpsi_6box},
centered on 3872~MeV, is the only bin with
a significant $B$ meson signal.  The $\DE$ distributions
(not shown) also have a significant $B$ meson signal
in the 3872~MeV bin, and only there.

The curves in the figures are the results
of binned likelihood fits that are applied simultaneously
to the $\Mbc$ and $\DE$ distributions.   The fit uses single Gaussians
for the $\Mbc$ and $\DE$ signals with widths fixed
at values determined from the $\chi_{c1}\rt\gamma\jp$ event sample.  
The areas of the $\Mbc$ and $\DE$ signal 
Gaussians are constrained to be equal.  
For the backgrounds, an ARGUS function~\cite{ARGUS} is 
used for $\Mbc$ and a first-order polynomial for $\DE$. 

\begin{figure}[htb]
\includegraphics[width=0.6\textwidth]{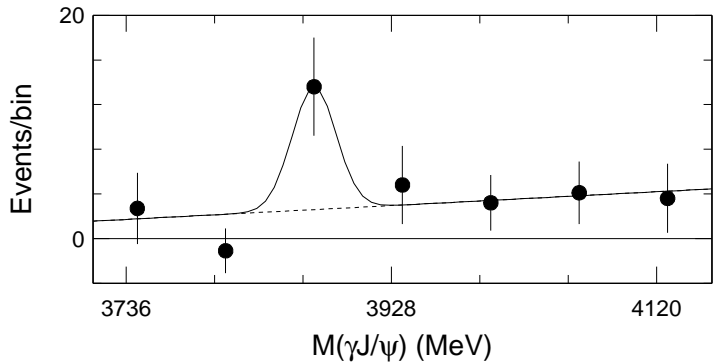}
\caption{ 
The signal yields from fits to the $\Mbc$ and $\DE$ distributions
in bins of $M(3872)$.  The curves are results of fits described in the text.
}
\label{fig:mgamjpsi_slicefit_64_1box}
\end{figure}

The results of the fits are plotted 
in Fig.~\ref{fig:mgamjpsi_slicefit_64_1box}.
The signal yield in the 3872~MeV bin is
$13.6\pm 4.4$ events.  We estimate 
the statistical significance
of the excess signal in the $M=3872$~MeV bin from
$\sqrt{-2\ln[{\mathcal L}(n_{\rm bkg})/{\mathcal L}(n_{\rm max})]}$,
where ${\mathcal L}(n)$ is the likelihood
value for a signal of $n$ events,
$n_{\rm bkg}$ is the background level
in the $X(3872)$ bin determined from  a linear 
fit to the event yields in the other bins
($2.6\pm 0.6$ events) inflated by $1\sigma$, and
$n_{\rm max}$ is the best-fit signal yield.
The significance is $4.0\sigma$.   

Using a MC-determined efficiency of $(19\pm 1)\%$,
we determine the  product branching fraction 
\begin{equation}
{\cal B}(B\rt K X){\cal B}(X\rt\gamma\jp)
 = 
(1.8 \pm 0.6 \pm 0.1 ) \times 10^{-6},
\label{eqn:br_gamm_jp}
\end{equation}
where the first error is statistical and the second systematic.
The latter includes effects of
differences between data and MC  resolutions
for $\Mbc$ and $\DE$ (3\%), tracking efficiency (3\%),
charged kaon identification (1\%), $K_S$ detection (3\%) 
and  $\gamma$ detection efficiency (2\%) added in
quadrature.  We assume that charged and neutral $B$
mesons have the same branching fraction to $K X(3872)$.
The product branching fraction of Eqn.~\ref{eqn:br_gamm_jp}
corresponds to the partial width ratio 
\begin{equation}
\Gamma(X\rt\gamma\jp)/\Gamma(X\rt\pipi\jp) = 0.14\pm 0.05.
\end{equation}

For the $X\rt\pipi\piz\jp$ search we select events with
a $\jp$, a kaon, and a $\pipi\piz$ combination.  The $\jp$,
kaon, $R_2$ and $\theta_B$  requirements are the same as for 
the $\gamma\jp$ search described above. For charged pions, 
we use any track that is not positively identified as a lepton.
Neutral pions are reconstructed from pairs of gamma rays detected in
the CsI calorimeter that fit a $\pi^0\rt\gamma\gamma$ hypothesis
with a $\chi^2<6$.  We require the $\pi^0$ cms momentum to be
greater than 180~MeV.  We eliminate events of the type
$B\rt K\pi^0\psip$; $\psip\rt\pipi\jp$  by vetoing events
where $M(\pipi\jp)$ is within $3\sigma$ of $m_{\psip}$.

In the case of
multiple entries corresponding to
different $\pi^0\rt\gamma\gamma$ assignments,
we select the $\gamma\gamma$ combination
with the best $\chi^2$ value for the $\pi^0\rt\gamma\gamma$ 
hypothesis.  For multiple entries
caused by different charged particle
assignments we select the track with the smallest impact parameter
relative to the $\jp\rt\leplep$ vertex.
The signal region is defined as $5.2725~{\rm GeV} < \Mbc <5.2875$~GeV
and $|\DE |< $ 0.035 GeV.
To select possible
$X(3872)\rt\pipi\piz\jp$ events we require 
$|\Delta M|<0.0165$~GeV, where
$\Delta M \equiv M(3\pi\jp)-3.872~{\rm GeV}$;
this corresponds to a $\pm 3\sigma$ requirement.

\begin{figure}[htb]
\includegraphics[width=0.6\textwidth]{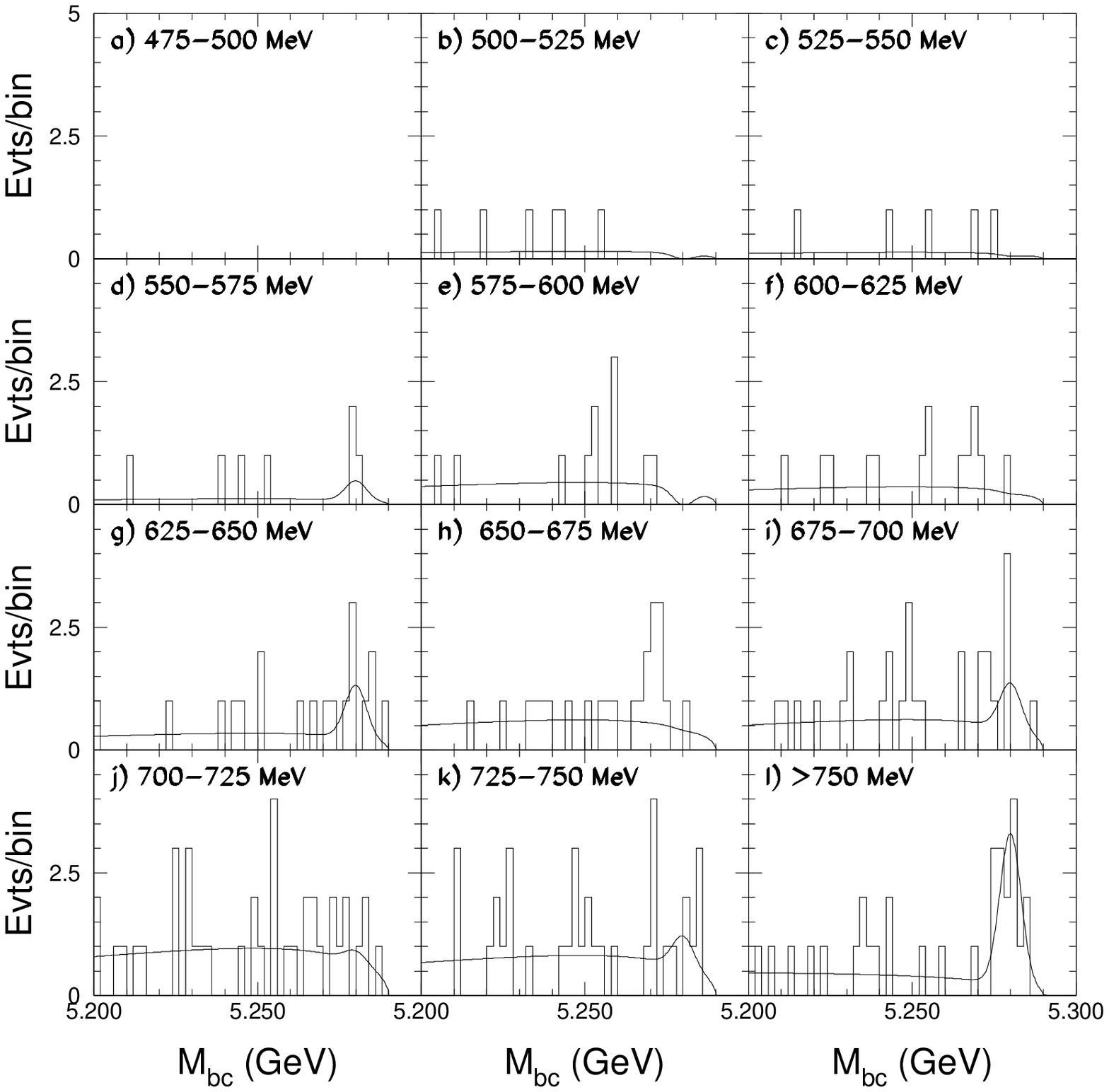}
\caption{
$\Mbc$ distributions for $B^-\rt K^-\pipi\piz\jp$
candidates in the $\DE$ and $X\rt\pipi\piz\jp$
signal regions for 25~MeV-wide $\pipi\piz$
invariant mass bins. 
}
\label{fig:mbc_12box}
\end{figure}

Figures~\ref{fig:mbc_12box}(a) through (l) show the $\Mbc$ 
distributions
for events in the $\DE$ and $X\rt\pipi\piz\jp$
signal regions for 25~MeV-wide  $\pipi\piz$
invariant mass bins.  
There is a distinct $B$ meson
signal for the $M(\pipi\piz)>750$~MeV bin
(Fig.~\ref{fig:mbc_12box}(l)),  and
no evident signals for any of the other $3\pi$ mass bins. 
The $\DE$ distributions
(not shown) exhibit a similar pattern.

The curves in the figures are the results
of binned likelihood fits that are applied simultaneously
to the $\Mbc$ and $\DE$ distributions.   The fit uses single Gaussians
for the $\Mbc$ and $\DE$ signals with widths fixed
at their MC-determined values.  The areas of the $\Mbc$ and $\DE$ signal 
Gaussians are constrained to be equal.  
For the backgrounds, an ARGUS function~\cite{ARGUS} is 
used for $\Mbc$ and a second-order polynomial for $\DE$. 

\begin{figure}[htb]
\includegraphics[width=0.6\textwidth]{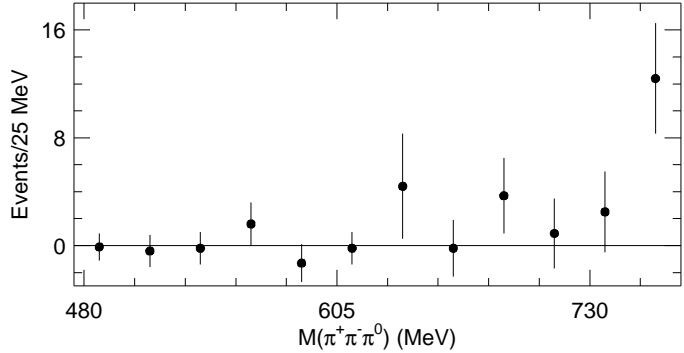}
\caption{ The $B$-meson signal
yield in bins of $3\pi$ invariant mass determined 
from  fits to the
$\Mbc$-$\DE$ distributions.}
\label{fig:m3pi}
\end{figure}

Figure~\ref{fig:m3pi} shows the fitted $B$-meson 
signal yields $vs$ $M(\pipi\pi^0)$.
All of the fitted yields are consistent
with zero except for the $M(\pipi\piz)>{\rm 750}$~MeV
bin, where the fit gives $12.4\pm 4.1$ events.

As possible backgrounds to the observed signal, we 
considered feed-across from $B\rt K\omega\jp$ decays and
non-resonant $B^-\rt K^-\pipi\piz\jp$ decays.


Belle has reported the observation of a near-threshold $\omega\jp$ 
mass enhancement in $B\rt K\omega\jp$ decays~\cite{skchoi_y3940}.
Although the peak of the $\omega\rt\pipi\piz$ resonance at $782.5$~MeV 
is 7.5~MeV above the maximum possible $3\pi$ invariant mass value
for $X\rt\pipi\piz\jp$ decays, there is some overlap
in the tails of the kinematically allowed regions
for the two processes that might result in some 
events from one signal feeding into the other.
From the fit results to the $\omega\jp$ mass
spectrum for $B\rt K\omega\jp$ of ref.~\cite{skchoi_y3940}
we determine that this source contributes
$0.75\pm0.14$~events to the overlap region.
As an independent check, we determined  the  $X(3872)\rt\pipi\piz\jp$
signal yield with a  more severe
restriction on $M(\pipi\piz\jp)$, namely
$M_X - 3\sigma < M(\pipi\piz\jp) < M_X + 1\sigma$,  that has
{\em no overlap} with the $\omega$ band.
The $X\rt\pipi\piz\jp$ signal yield in the truncated region is 
$10.6\pm 3.6$ events.  For a Gaussian signal distribution with
no feed-across background, we expect the truncation of the signal region
to reduce the signal by 2.1~events (16\%); the observed reduction of 
1.8~events is consistent with a feed-across level that
is less than one event.

To study possible contributions to the $X(3872)\rt\pipi\piz\jp$
signal from non-resonant $B^-\rt K^-\pipi\piz\jp$ 
decays, we defined two side-regions in the 
[$\Delta M, M(3\pi)$] plane.  
The ``upper'' side-band is 
centered at $\Delta M = 0.033$~GeV 
(just above the $X(3872)$ signal region)
and is $\pm 0.0165$~GeV wide, with
$660~{\rm MeV}<M(3\pi)<{\rm 760}$~MeV.
The ``lower'' sideband is $\pm 0.0165$~GeV wide 
centered at $\Delta M = -0.033$~GeV with $M(3\pi)>640$~MeV. 
Each sideband has an area that is four times that of the 
$X\rt\pipi\piz\jp$ signal bin. 

There is no evidence for significant signal yields in
either sideband.  Fits to the $\Mbc$-$\DE$ distributions 
give $4.3\pm 6.2$ and $6.4\pm 5.6$ signal events
for the upper and lower bands, respectively.  
Normalizing by area, this gives an estimate of a
possible non-resonant background level
in the $X\rt\pipi\piz\jp$ signal 
bin of $1.3\pm 1.0$ events.

The combined estimate for feed-across and non-resonant
backgrounds is $2.1 \pm 1.0$.  We inflate this by $+1\sigma$ 
and compute the significance 
with $n_{\rm bkg}=3.1$~events and
$n_{\rm max}=12.4$~events.
The resulting statistical significance is
$4.3\sigma$.

To determine the branching fraction for $B\rt\pipi\piz\jp$
we attribute all of the signal events 
with $M(\pipi\piz)>{\rm 750}$~MeV to $X\rt\pipi\piz\jp$ decay. We
compute the ratio of $\pipi\piz\jp$ and $\pipi\jp$ branching
fractions by comparing this to the number of  $X\rt\pipi\jp$
events in the same data sample, corrected by MC-determined 
relative detection efficiencies.  
The ratio of branching fractions is 
\begin{equation}
\frac{{\cal B}(X\rt\pipi\piz\jp)}{{\cal B}(X\rt\pipi\jp)}
= 1.0\pm0.4{\rm (stat)}\pm 0.3 {\rm (syst)},
\end{equation}
where the systematic error reflects the uncertainty in the 
relative acceptance, the level of possible feed-across and
nonresonant background, and possible event loss
due to the $M(3\pi)>750$~MeV requirement, all added in quadrature.  

In summary, we report strong evidence for the 
decays $X(3872)\rt\gamma\jp$
and  $X(3872)\rt\pipi\piz\jp$.
These are the first measurements of an $X(3872)$ decay mode other than $\pipi\jp$.
The $X(3872)\rt\gamma\jp$ result unambiguously establishes
the charge conjugation parity of the $X$ as $C=+1$.
This, in turn, would mean that the $\pipi$ system in 
$X\rt\pipi\jp$ decay is from the decay of a $\rho$
meson, as is suggested by the observed $\pipi$ invariant
mass distribution

In the $X\rt 3\pi\jp$ signal, the
$\pipi\piz$ invariant masses are clustered above 750~MeV,
near the upper kinematic boundary.  This is suggestive of a 
sub-threshold decay to a virtual $\omega$ plus a $\jp$, 
and is further evidence for $C=+1$.  This decay mode
was predicted by Swanson to occur at about 
the observed strength based on a model where the $X(3872)$ 
is  considered to be primarily a $D^0\bar{D}^{0*}$ hadronic 
resonance~\cite{swanson_1}.  

The large isospin violation implied by the near equality of
the $\rho\jp$ and $\omega\jp$ decay widths is difficult to accomodate
in a $c\bar{c}$ charmonium interpretation of the $X(3872)$, but a natural
consequence of the meson-meson bound state model point of 
view~\cite{tornqvist,swanson_1}.

We thank the KEKB group for the excellent operation of the
accelerator, the KEK Cryogenics group for the efficient
operation of the solenoid, and the KEK computer group and
the National Institute of Informatics for valuable computing
and Super-SINET network support. We acknowledge support from
the Ministry of Education, Culture, Sports, Science, and
Technology of Japan and the Japan Society for the Promotion
of Science; the Australian Research Council and the
Australian Department of Education, Science and Training;
the National Science Foundation of China under contract
No.~10175071; the Department of Science and Technology of
India; the BK21 program of the Ministry of Education of
Korea and the CHEP SRC program of the Korea Science and
Engineering Foundation; the Polish State Committee for
Scientific Research under contract No.~2P03B 01324; the
Ministry of Science and Technology of the Russian
Federation; the Ministry of Education, Science and Sport of
the Republic of Slovenia; the National Science Council and
the Ministry of Education of Taiwan; and the U.S.\
Department of Energy.


\begin{thebibliography}{99}
\bibitem{conj} In this paper the inclusion of charge-conjugate modes
is always implied.
\bibitem{X_mass} This is the weighted average of all of the reported $X(3872)$
mass measurements.
\bibitem{m3pijpsi}
In the analyses reported here we 
require $|M(\leplep)-M_{\jp}|<0.02$~GeV and reduce
lepton momentum resolution effects
by defining 
$M(\gamma\jp)\equiv
M(\gamma\leplep) - M(\leplep) + M_{\jp}$
(and 
$M(\gamma\jp)\equiv
M(\pipi\pi^0\leplep) - M(\leplep) + M_{\jp}$), where
$M_{\jp}=3.0969$~GeV is the Particle Data Group~\cite{PDG} value
for the $\jp$ mass.
\end{thebibliography}
\end{document}